\documentclass[a4paper,12pt]{article}
\usepackage{graphicx}
\usepackage{geometry}
\usepackage{amsmath,amssymb}
\usepackage{times}
\usepackage{indentfirst}
\usepackage{hyperref}
\usepackage{algpseudocode}
\usepackage{algorithmicx,algorithm}
\hypersetup{hypertex=true,colorlinks=true,linkcolor=blue,anchorcolor=blue,citecolor=blue}

\title{\bf Simple rejection Monte Carlo algorithm and its application to multivariate statistical inference}
\author{Fengyu Li$^{1,2}$, Huijiao Yu$^{1,2}$, Jun Yan$^{1,2}$, Xianyong Meng$^{*1,2}$\\
\small$^1$ Department of Applied Mathematics, School of Information Science and Engineering,\\\small Shandong Agricultural University, Taian, Shandong\\
\small$^2$ Huanghuaihai Key Laboratory of Smart Agricultural Technology, Ministry of Agriculture and Rural Affairs,\\\small Taian, Shandong}
\date{}

\begin{document}

\maketitle

\begin{abstract}
The Monte Carlo algorithm is increasingly utilized, with its central step involving computer-based random sampling from stochastic models. While both Markov Chain Monte Carlo (MCMC) and Reject Monte Carlo serve as sampling methods, the latter finds fewer applications compared to the former. Hence, this paper initially provides a concise introduction to the theory of the Reject Monte Carlo algorithm and its implementation techniques, aiming to enhance conceptual understanding and program implementation. Subsequently, a simplified rejection Monte Carlo algorithm is formulated. Furthermore, by considering multivariate distribution sampling and multivariate integration as examples, this study explores the specific application of the algorithm in statistical inference.

\end{abstract}

\section{Introduction}\label{se:1}
Statistical inference plays a crucial role in all scientific fields, serving to provide theoretical underpinnings for quantitative models of uncertainty and validate data-based models. Modern statistical inference methods based on sampling enable us to comprehend complex phenomena (Warne et al., 2018). Sampling can be categorized into two types: population sampling and data set sampling. The former is often referred to as Monte Carlo or random simulation, which can be further divided into distribution function-based sampling (Meng et al., 2022) and density function-based sampling (Chen and Yang, 2006; Andrieu and Freitas, 2003). The latter is commonly known as subsampling (Ai et al., 2021; Li et al., 2022), which also falls under the category of Monte Carlo when we consider a data set as a special representation of the population.

Monte Carlo sampling is commonly employed for generating random numbers of random variables or random vectors, which are then utilized to estimate unknown mathematical expectations and their confidence intervals. This well-known algorithm, known as the Monte Carlo algorithm, can also provide an estimation of the complete distribution (L’Ecuyer et al., 2022). The fundamental concept behind the Monte Carlo algorithm lies in utilizing sampled data to perform various calculations and feature analyses. Generating random numbers that adhere to a probability distribution model becomes a crucial technology for these algorithms, often accomplished by computers. There exist numerous options for sampling methods; among them, Markov Chain Monte Carlo (MCMC) (Robert and Casella, 2010) is widely adopted due to its ease of implementation. MCMC technique serves as the gold standard technique for Bayesian inference (Nemeth and Fearnhead, 2021), frequently used to extract samples from challenging posterior distributions (Vyner et al., 2023). However, correlations exist between data generated by MCMC, posing challenges in convergence assessment. In recent years, the Rejection Monte Carlo (RMC) algorithm has emerged sporadically in literature under alternative names such as screening sampling or reject-accept sampling algorithms (Warne et al., 2018; Martino and M´ıguez, 2011; Ma et al., 2022).

Compared to MCMC technology, there is a relatively limited amount of literature on RMC technology. RMC is commonly employed for constructing complex Monte Carlo algorithms, but it itself serves as a standard Monte Carlo technique that filters samples generated by the proposal distribution through the density function to obtain samples from the target distribution (Chen and Yang, 2006; Maryino and M´ıguez, 2011). In comparison with MCMC, RMC possesses unique advantages (Warne et al., 2018): (1) RMC is user-friendly; (2) The generated samples are independent and equally distributed; (3) Calculation efficiency remains unaffected by calculation parameters, particularly without requiring heuristic definition of transition probability. However, the drawbacks of RMC are also evident as it can be challenging to determine an appropriate recommendation sampling distribution. When the recommendation distribution significantly deviates from the target distribution, computational efficiency becomes low. Therefore, previous studies have primarily focused on sampling efficiency and determining efficient recommendation distributions (Chen and Yang, 2006; Maryino and M´ıguez, 2011). Nevertheless, with continuous advancements in computer technology, efficiency no longer poses a key constraint in many cases. The flexibility of RMC technology towards suggestion distributions has been further enhanced thereby making its application more widespread. Consequently,this paper provides enlightening insights into the convenient learning perspectives and program implementation aspects of RMC techniques to enable arbitrary model sampling using RMC methods—especially for multivariate models—and explores their versatility and practical value.
\section{Theoretical Foundation}\label{se:2}
The RMC algorithms are a versatile approach for generating independent samples from a target density (Martino and M´ıguez, 2011). Let $f(x)$ be the probability density function of a target random variable $X$, and let $f_0(x)$ the suggested density function that can be easily sampled, with the same support supp(f) as the function $f(x)$. Additionally, let $c$ a constant satisfying the following condition:
\begin{equation}c{f_0}(x) \geqslant f(x).\end{equation}

The RMC algorithms are typically delineated in the literature as follows.
\begin{algorithm}[h]
\caption{General RMC algorithm}
\label{alg:1}
\begin{algorithmic}[1]
\State  Let $i=1$.
\State Generate a random number $x$ from ${f_0}(x)$, and a random number $y$ from uniform distribution $U[0,1]$.
\State If $\frac{{f({{\rm x}})}}{{c{f_0}(x)}} \geqslant y$, then ${x_i} = x$, $i = i + 1$; otherwise, reject $x$ and run step 2.
\State If $i\geq N$, the procedure is discontinued; otherwise step 2 is run.
\end{algorithmic}
\end{algorithm}
A merit function, such as $cf_0(x)$ in Inequality $(1)$ plays a crucial role in the efficiency of the sampling. When the deviation between the proposed model ${f_0}(x)$ and $f(x)$ is significant and  $c$ is chosen excessively large, the difference  between the resulting merit function and objective function, $cf_0(x)-f(x)$, becomes substantial. This leads to a low acceptance probability $P\{ y \leqslant \frac{{f(x)}}{{c{f_0}(x)}}\}$, causing most points to be rejected during  random selection and ultimately reducing sampling efficiency.

The construction of $f_0(x)$ is typically designed to enhance sampling efficiency by closely approximating $f(x)$, while the constant $c$ is determined as the $c = \arg {\min _{\text{d}}}\{ d{f_0}(x) \geqslant f(x),x \in {\text{supp}}(f)\}$.

{\bf Theorem 1} Let the value $x$ of random variable X be a random number generated by the Algorithm $1$, then X $\sim f(x)$.

{\bf Proof} The density functions of random numbers $x$ and $y$ are  denoted as ${f_0}(x)$ and ${f_1}(y) = \left\{ {\begin{array}{*{20}{c}}
  {1,}&{y \in [0,1]} \\
  {0,}&{\rm other.}
\end{array}} \right.$, respectively, according to the Algorithm \ref{alg:1}. As the processes generating $x$ and $y$ are mutually independent, the joint probability density function of $x$ and $y$ can be expressed as follows:
\begin{equation}f(x,y) = {f_0}(x){f_1}(y).\end{equation}

The cumulative probability function of random number $x$:
$$\begin{aligned}
  P\{ x \leqslant c\}  &= P\{ x \leqslant c|y \leqslant \frac{{f(x)}}{{c{f_0}(x)}}\}  = \frac{{P\{ x \leqslant c,y \leqslant \frac{{f(x)}}{{c{f_0}(x)}}\} }}{{P\{ y \leqslant \frac{{f(x)}}{{c{f_0}(x)}}\} }} = \frac{{\int_{- \infty}^c {\int_0^{\frac{{f(x)}}{{c{f_0}(x)}}}{f(x,y){\rm d}x{\rm d}y}}}}{{\frac{{f(x)}}{{c{f_0}(x)}}}} \\
&= \int_{ - \infty }^c {f(x){\rm d}x}.\#\\
\end{aligned}.$$
The identification of the optimal merit function constitutes a fundamental aspect in existing RMC literature (Chen and Yang, 2006; Martino and M\'{i}gues, 2011). However, its practical implementation poses significant challenges, thereby becoming a bottleneck that hinders the progress of the RMC algorithm. According to Algorithm 1, the construction of the proposed distribution is unrestricted and may deviate from the target distribution. This discrepancy inevitably leads to inefficient sampling; nevertheless, advancements in computer technology have compensated for this limitation. Consequently, contemporary approaches favor either uniform distribution models or segmented uniform distribution models as suitable candidates for the proposed distributions. By employing uniformly distributed or segmented uniformly distributed merit functions, it is possible to significantly enhance both versatility and comprehensibility within RMC techniques while effortlessly facilitating computer-based sampling across arbitrary distribution models.

\begin{figure}[!ht]
\centering
\includegraphics[width=0.6\linewidth]{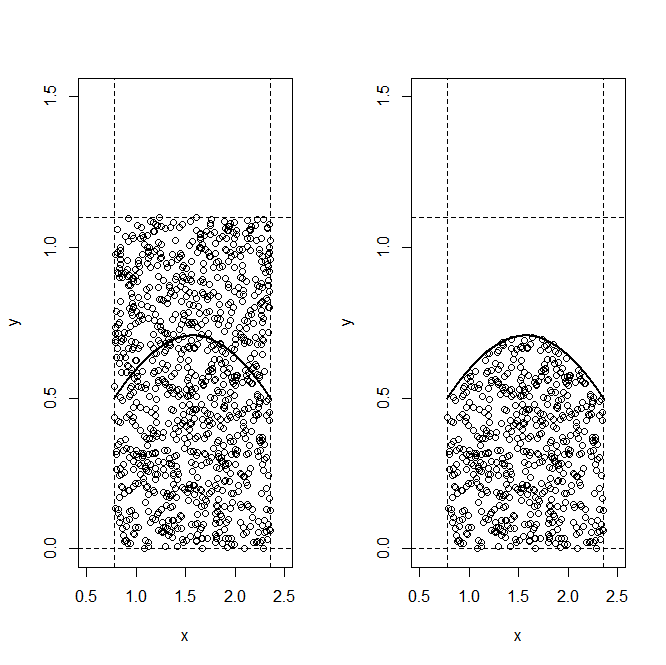}
\caption{RMC sampling principle demonstration diagram}
\label{fig:1}
\end{figure}

An intuitive understanding of sampling from the target density is to allocate more samples in regions with higher density values and fewer samples in regions with lower density values, proportionally matching the value of the density function. This can be easily achieved by employing a uniform distribution. For instance, when considering sampling from a given density function
 \begin{equation}f(x) = \sin x/\sqrt 2 ,x \in (\pi/4,3\pi/4).\end{equation} The uniform distribution can be employed to evenly distribute points across the rectangular area $[\pi/4,3\pi/4]\times[0,1.1]$. Subsequently, the points located between the density function and the x-axis effectively reflect the characteristics of the density function and are thus considered valid. As depicted in Figure \ref{fig:1}, these accepted points correspond to samples based on their horizontal coordinates.

The previous analysis suggests using a simple technique called Simple Rejection Monte Carlo (SRMC) to sample from a model $f(x)$, which can be described as Algorithm 2.

\begin{algorithm}[h]
\caption{SRMC algorithm}
\label{alg:2}
\begin{algorithmic}[1]
\State Let $i = 1$;
\State Draw evenly a random number $x$ from the support interval of $f(x)$, and a random number $y$ from the interval $[0,c](c \geqslant \max f(x),x \in [a,b])$
\State If $f(x) > y$, then ${x_i} = x$, $i = i + 1$; otherwise, reject $x$ and run step 2.
\State If $i > N$, the procedure is discontinued; otherwise step 2 is run.
\end{algorithmic}
\end{algorithm}

Given the ease of drawing samples uniformly from the support set  ${\rm suff}(f)$ of function $f(x)$, Algorithm 2 can be straightforwardly extended to a Simple Rejection Monte Carlo (SRMC) technique for sampling from a multivariate random variable model  $f({x_1},{x_2}, \cdots ,{x_d})$, as described in algorithm 3.

\begin{algorithm}[h]
\caption{SRMC algorithm for random vectors}
\label{alg:3}
\begin{algorithmic}[1]
\State Let $i = 1$;
\State Extract uniformly a random array $x^0=(x_1^0, x_2^0,\cdots, x_d^0)$ from the support ${\rm suff}(f)$ of the multivariate model $f(x_1, x_2, \cdots, x_d)$ and a random number $y$ from the interval $[0,c](c \geqslant \max f(x_1, x_2, \cdots, x_d),(x_1, x_2, \cdots, x_d) \in {\rm suff}(f))$;
\State If $f(x^0)>y$, then $x^i =x^0$, $i = i + 1$; otherwise, reject $x^0$ and run step 2
\State If $i>N$, the procedure is discontinued; otherwise step 2 is run.
\end{algorithmic}
\end{algorithm}

\section{Sampling implementation of multivariate distribution models}\label{se:3}
As computing technology continues to advance, computational efficiency is often no longer the primary constraint, and the implementability of algorithms assumes greater significance. Algorithm \ref{alg:2} and Algorithm \ref{alg:3} can be easily implemented programmatically, thereby facilitating intelligent computation. The evaluation of algorithm strengths and weaknesses can be measured in terms of accuracy and program runtime. The following specific examples demonstrate that RMC out-sampling for multivariate distributions can be readily implemented, while considering bias and program runtime as indicators of accuracy and feasibility. All programs are implemented using the R language (Meng, 2022).

{\bf Example 1} Let the density function of a two-dimensional Gaussian random vector $(X,Y)$ be
\begin{equation}f(x,y) = \frac{1}{{2\pi \sqrt{\left| \sum  \right|}}}\exp \{-\frac{1}{2}{\mu ^T}{\sum}^{-1}\mu\}\end{equation}
where $\mu  = \left( {\begin{array}{*{20}{c}}
  x \\
  y
\end{array}} \right) \in {R^2}$, $\sum  = \left( {\begin{array}{*{20}{c}}
  1&{0.2} \\
  {0.2}&1
\end{array}} \right)$. Please generate a sample of the random vector by RMC technology.

\begin{figure}[!ht]
\centering
\includegraphics[width=0.8\linewidth]{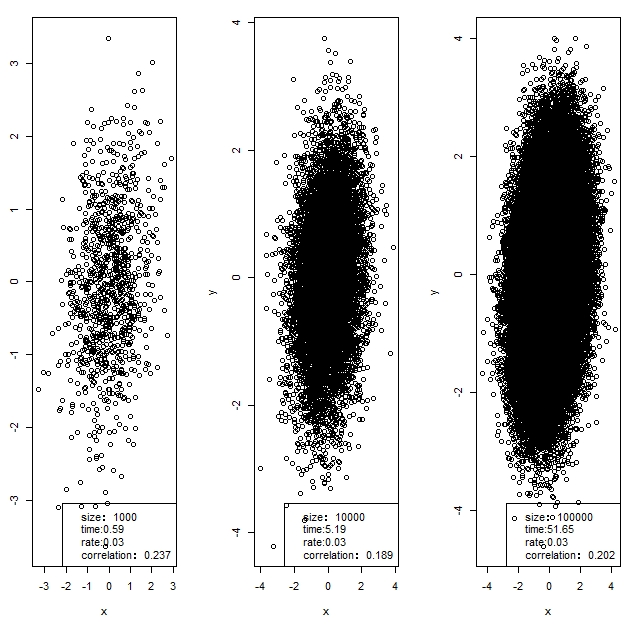}
\caption{Scatterplot of RMC sampling with binary Gaussian distribution}
\label{fig:2}
\end{figure}

According to Algorithm 3, the square region  $[-5,5]\times [-5,5]$ is selected with  $c=\max f(x,y)=0.1657$. The scatter plot of the sampling is presented in Figure \ref{fig:2}. The sample sizes from left to right are: 1000, 10000 and 100000; corresponding program running times are 0.59, 5.19 and 51.65 seconds respectively. The correlation coefficients calculated from the extracted samples are found to be 0.237, 0.189 and 0.202 respectively, which exhibit slight deviation from the theoretical correlation coefficient of 0.2; moreover, as the number of samples increases, the absolute value of this deviation decreases accordingly.The scatterplot (Figure \ref{fig:2}) visually demonstrates the structural relationship between variables where most values fall within the square region $[-4, 4]\times [-4, 4]$, while occurrences outside this range can be considered negligible.

Probability calculations play a crucial role in statistical inference, often necessitating the use of integral computations. However, solving these integrals analytically can be challenging, thus making sampling methods an invaluable tool for their resolution. This is exemplified below through the application of ordinary dual integration.

{\bf Example 2} Calculate the dual integral $\iint_{{\rm D}} {xy{{\rm d}}x{{\rm d}}y}$, where the region D is bounded by ${y^2} = x$, $x$ axis and $y = x - 2$, and $D=\{(x, y)\mid y\geq 0\}.$

Construct a rectangular area $S = [0,4] \times [0,2] \supseteq D$. Then the integrated function $xy$ is easily integrated over $S$, and the function
 $$f(x,y) = \left\{ {\begin{array}{*{20}{c}}
  xy/\iint_{{\rm S}} {xy{{\rm d}}x{{\rm d}}y},(x,y) \in S \\
  0 ,(x,y) \notin S
\end{array}} \right.$$ can be viewed as a density function of a random vector $(X,Y)$. Then the required dual integral can be transformed into
$$\begin{aligned}
  \iint_{{\rm D}} {xy{{\rm d}}x{{\rm d}}y} &= \iint_S {xy{{\rm d}}x{{\rm d}}y}\iint_D {(xy/\iint_S {xy{{\rm d}}x{{\rm d}}y}){{\rm d}}x{{\rm d}}y} = \iint_S {xy{{\rm d}}x{{\rm d}}y}\iint_S {{\rm I}(x,y)f(x,y){{\rm d}}x{{\rm d}}y} \\
\end{aligned}$$
where ${\rm I}(x,y)$ is a representative function of the set D, i.e., when $(x,y)\in D$, ${\rm I}(x,y)=1$ ; otherwise, ${\rm I}(x,y)=0$. Since $f(x,y)$ is a density function on S, it follows that$\iint_SI(x,y)f(x,y)dxdy=E(I(X,Y))\approx\frac{{\#} \{(x,y)\mid(x,y) \in D\}}{{\# \{ (x,y)\mid(x,y) \in S\} }},$ where the symbol $\#$ denotes the number of elements in the corresponding set. If one considers a uniform distribution over the region S, then $\iint_S {xy{{\rm d}}x{{\rm d}}y}=8\iint_S xy\frac{1}{8}{\rm d}x{\rm d}y=8E(XY)\approx 8\frac{1}{N}\sum_{i=1}^Nx_iy_i.$ Therefor, the target integral can be approximated as follows.
\begin{equation}
  \iint_{{\rm D}} {xy{{\rm d}}x{{\rm d}}y} \approx 8\frac{1}{N}\sum\limits_{i = 1}^N {{x_i}{y_i}} \frac{{\#} \{(x,y)\mid(x,y) \in D\}}{{\# \{ (x,y)\mid(x,y) \in S\} }}.\\
\end{equation}
From Equation (5) and Algorithm \ref{alg:3}, in order to complete the computation of the objective integral, first a uniform sampling should be performed on the region S, and then the resulting uniform samples should be screened by the density function $f(x,y)$. Assume that the set of uniform random numbers drawn first is denoted as S1 and the set of random numbers obtained by subsequent screening is S2. Then the objective integral can be completed by calculating $\frac{1}{N}\sum\limits_{i = 1}^N {{x_i}{y_i}}$ and $\frac{{\#} \{(x,y)\mid(x,y) \in D\}}{{\# \{ (x,y)\mid(x,y) \in S\} }}$ from S1 and S2, respectively. The results were calculated using R software programming as follows, with 10 runs for each sample size, and the results are averaged.

\begin{table}[ht]
\centering
\label{tab:1}
\begin{tabular}{|l|l|l|l|l|l|}
\hline
Samples(S2) & 100 & 1000 & 10000 & 100000 &  1000000\\
\hline
Estimated value-real value & $-$0.173 & 0.032 & 0.059 & $-$0.009  & 0.00046\\
\hline
Run time (seconds) & 0.0062 & 0.055 & 0.56 & 5.49 &  50.1\\
\hline
\end{tabular}
\end{table}

\section{Summary}\label{se:4}
Previous literature primarily focuses on rejection sampling algorithms for one-dimensional random variables, which involve constructing a proposed sampling model and optimal function that are both easy to sample from and closely approximate the target model. However, despite its high sampling efficiency, the constructed proposed model lacks operability, limiting its intelligent implementation in statistical inference. This paper highlights the versatility and ease of operation provided by the RMC technique and introduces the SRMC technique as a solution. The proposed technique is applicable to both one-dimensional and multivariate random variables, offering generalizability and comprehensibility. Specific examples demonstrate that this algorithm can be easily programmed and implemented in multivariate statistical analysis with high accuracy. Consequently, it will undoubtedly facilitate widespread adoption of the SRMC technique while providing an alternative method for intelligent implementation of multivariate statistical inference.

\end{document}